\documentclass[conference]{IEEEtran}[anon]

\usepackage{listings}
\usepackage{tikz}
\usepackage{pgfplots}
\usepackage[caption=false,font=footnotesize]{subfig}

\usepackage[T1]{fontenc}
\usepackage[utf8]{inputenc}
\usepackage[most]{tcolorbox}
\usepackage{textcomp}
\usepackage{listings}
\usepackage{tikz}
\usetikzlibrary{arrows.meta,positioning,fit,calc,backgrounds}

\usepackage[T1]{fontenc}
\usepackage[utf8]{inputenc} 
\usepackage{microtype}    
\usepackage{graphicx}
\usepackage{listings}
\usepackage[most]{tcolorbox}
\usepackage{makecell}

\usepackage[table]{xcolor}
\usepackage{colortbl}
\usepackage{booktabs}

\usepackage{amssymb}

\definecolor{azure}{rgb}{0.0, 0.5, 1.0}
\definecolor{awesome}{rgb}{1.0, 0.13, 0.32}

\lstdefinestyle{hls}{
  language=C++,
basicstyle=\ttfamily\footnotesize,
keywordstyle=\ttfamily\footnotesize\color{azure},
stringstyle=\ttfamily\footnotesize\color{awesome},
  numbers=none,
  breaklines=true,
  breakatwhitespace=false,
  columns=fullflexible,
  keepspaces=true,
  tabsize=2,
  showstringspaces=false,
  upquote=true,
  directives=true
}

\begin{document}
%
\title{LAAFD: LLM-based Agents for Accelerated FPGA Design}

\author{
\IEEEauthorblockN{
Maxim Moraru,
Kamalavasan Kamalakkannan,
Jered Dominguez-Trujillo,\\
Patrick Diehl,
Atanu Barai,
Julien Loiseau,\\
Zachary Kent Baker,
Howard Pritchard,
Galen M.\ Shipman
}
\IEEEauthorblockA{
Los Alamos National Laboratory\\
Los Alamos, NM, USA
}
}

\maketitle

\begin{abstract}
FPGAs offer high performance, low latency, and energy efficiency for accelerated computing, yet adoption in scientific and edge settings is limited by the specialized hardware expertise required. High-level synthesis (HLS) boosts productivity over HDLs, but competitive designs still demand hardware-aware optimizations and careful dataflow design. We introduce LAAFD, an agentic workflow that uses large language models to translate general-purpose C\texttt{++} into optimized Vitis HLS kernels. LAAFD automates key transformations: deep pipelining, vectorization, and dataflow partitioning and closes the loop with HLS co-simulation and synthesis feedback to verify correctness while iteratively improving execution time in cycles. Over a suite of 15 kernels representing common compute patterns in HPC, LAFFD achieves 99.9\% geomean performance when compared to the hand tuned baseline for Vitis HLS. For stencil workloads, LAAFD matches the performance of SODA, a state-of-the-art DSL-based HLS code generator for stencil solvers, while yielding more readable kernels. These results suggest LAAFD substantially lowers the expertise barrier to FPGA acceleration without sacrificing efficiency.
\end{abstract}


%
\IEEEpeerreviewmaketitle

\section{Introduction}
Field Programmable Gate Arrays (FPGAs) are ubiquitous for tasks ranging from low level interfacing with sensors and digital signal processing to accelerating Large Language Model (LLM) inference. The main hurdles are longer development cycles and the specialized expertise needed to implement applications and algorithms on FPGAs. Although high-level synthesis (HLS) tools let developers describe designs in C/C\texttt{++}-like languages (C\texttt{++} for Vitis HLS for AMD FPGAs and SYCL for Altera FPGAs), achieving efficient hardware still requires careful customization (\emph{e.g.,} pragmas, memory architecture choices, and a dataflow programming model) to produce optimized implementations.


Prior work shows that LLMs can translate code across languages and programming models. However, mapping a software-level algorithm into a hardware-oriented high-level description (\emph{e.g.,} for HLS) is substantially more complex and error-prone, often producing suboptimal implementations. The FPGA design space is far larger than software targeting fixed hardware. For these reasons, a naïve prompt to translate a C\texttt{++} kernel directly into HLS C\texttt{++} is unlikely to be efficient.

Agentic workflows powered by large language models offer a promising way to automate parts of the HLS development process.  Major vendors’ HLS tools report kernel latency (in cycles) and resource utilization, and support verification at both the C\texttt{++} and RTL levels. These reports and simulation results can then guide the agents to refine the translated kernels and confirm their correctness.

In this work we develop a set of kernels that requires several HLS specific optimizations such as pipelining, vectorization, dataflow, and data reuse using a custom memory hierarchy. We utilize these kernels along with test benches to validate the LLM-generated C\texttt{++} kernels for Vitis HLS and to optimize them for the latency. The agentic workflow developed in this paper demonstrates the ability to apply state of the art HLS optimizations for this set of kernels, which includes several stencil solvers. All the kernels tested achieved near performance parity with manually optimized HLS implementations. 

Specifically, we make the following contributions towards automatically translating C\texttt{++} kernels to performance optimized FPGA HLS kernels:
\begin{itemize}
    \item \textit{LAAFD}, an agentic workflow that utilizes feedback from HLS reports and C\texttt{++} co-simulation to generate latency optimized C\texttt{++} for HLS implementation.
    \item A comparison of the optimization capabilities of the latest LLMs such as gpt-5, gpt-5-nano and o4-mini for LAAFD.
    \item Performance evaluation of the optimized HLS kernels generated by this agentic workflow compared to the optimized hand-written kernels or to a state of the art domain specific framework.
\end{itemize}







\section{Background}
Many frameworks, domain specific languages, and approaches attempt to bridge the gap between algorithmic and software implementation to optimized and high-performance hardware implementations. This sections gives a basic background on several of these approaches, focusing on high-level synthesis and RTL generation frameworks.




\subsection{High level synthesis (HLS)}
\label{sec:HLS}

FPGAs have traditionally been programmed with HDLs such as \textit{verilog} and \textit{VHDL}, demanding expertise in digital circuit design and long development and verification cycles to achieve an optimal design. High-level synthesis tools increase productivity by letting developers describe architectures in C/C\texttt{++} (e.g. for AMD FPGAs) and SYCL (e.g., for Altera FPGAs). HLS can automatically apply several loop optimizations when guided by \texttt{\#pragma hls}, such as vectorization and pipelining, but more sophisticated optimizations (complex data reuse, tiling, dataflow parallelism) still require manual code restructuring and custom data movement logic. Consequently, converting sequential code or code written for shared/distributed models into an efficient FPGA implementation remains a labor-intensive task.


\begin{figure}[t]
  \centering
  \footnotesize
    \includegraphics[width=0.49\textwidth]{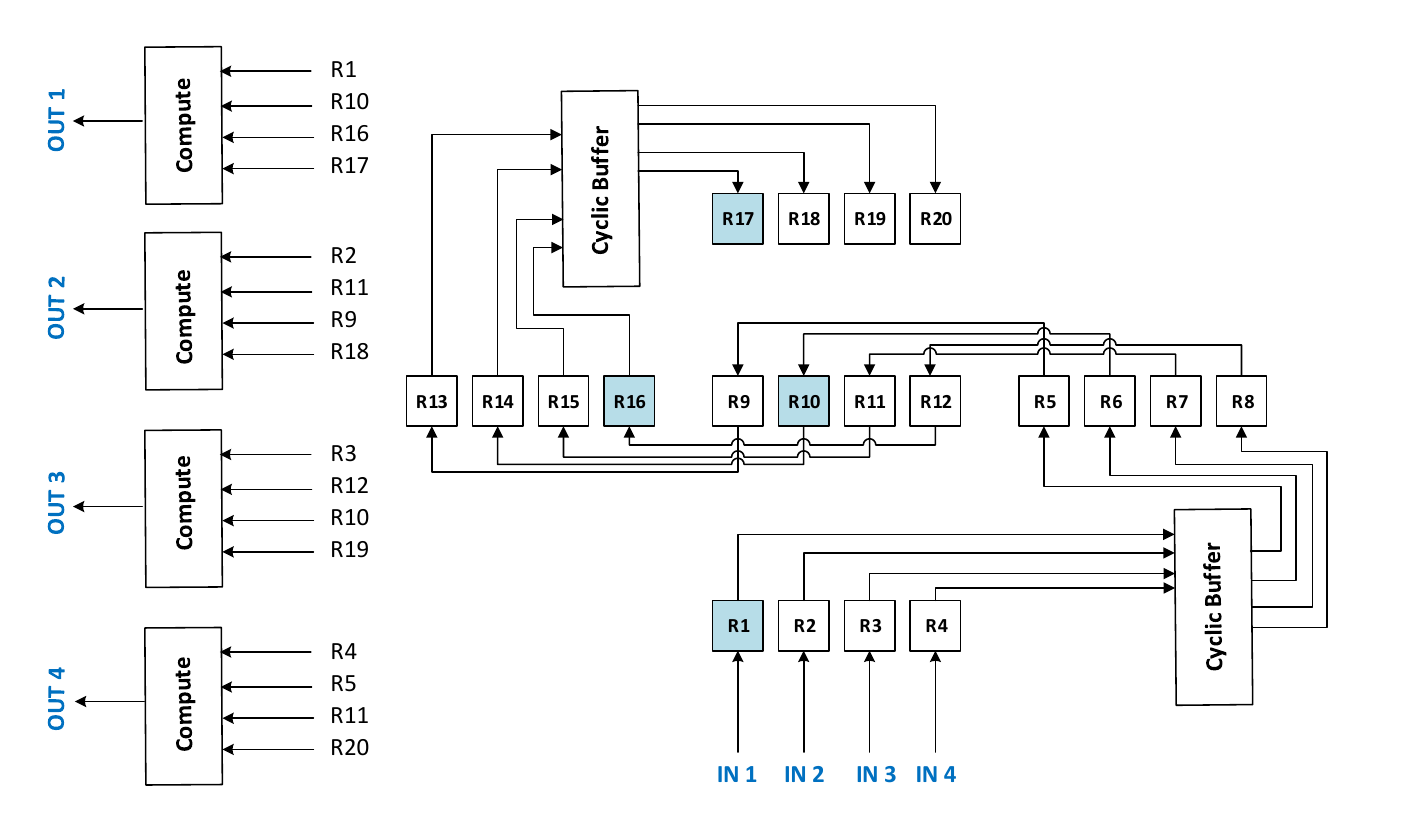}
  \caption{Dataflow in compute loop of stencil 2D (S2D) iteration}
  \label{fig:accel}
\end{figure}

State-of-the-art optimization of a 5-point 2D stencil kernel demands extensive HLS customizations before synthesis to obtain the dataflow architecture shown in Figure~\ref{fig:accel}. Typically, the computation and memory access units are split into separate modules that communicate via streams, and each module must be annotated with pragmas that force parallel execution of the merged regions. The compute function must be adapted to read from the stream, buffer data using minimal on-chip memory, and support wider data paths (e.g., vectorization) and multi-iteration parallelism, which requires feeding the output of one compute instance into the next. Moreover, the kernel interface must expose parameters such as the maximum number of outstanding transactions and burst size. Consequently, a compact C++ kernel of roughly ten lines can expand into a highly tuned HLS implementation spanning hundreds of lines.


High-level synthesis tools such as Vitis HLS convert C/C\texttt{++} kernels into dataflow architectures by analyzing the source structure and applying user-provided pragma hints. Directives like \texttt{pragma HLS pipeline} and \texttt{pragma HLS dataflow} cause the compiler to pipeline loop iterations and instantiate a dataflow architecture, while \texttt{HLS tripcount} lets the tool estimate loop latency when iteration counts are known. In addition to generating optimized RTL, HLS tools produce synthesis reports that enumerate FPGA resource usage (LUTs, registers, BRAMs, DSPs) and report latencies for individual loops, modules, and whole kernels. This information lets designers evaluate the quality of the generated implementation and balance performance, power and, area.

HLS accurately predicts loop latency when trip counts are supplied, but it does not model the inter-loop latency introduced by a \texttt{DATAFLOW} region. As a result, end-to-end latency can be slightly under-reported for multi-stage dataflow designs, which explains why, in Tables \ref{tab:Kernels} and \ref{tab:Soda_Kernels}, the ``ideal" latency sometimes exceeds the tool-reported latency. Nonetheless, the reported cycle count remains a consistent metric for comparing kernel performance within our workflow.

Finally, HLS supports co-simulations of the original C++ code with either the synthesizes C++ model or the generated RTL, providing essential validation of functional correctness.

\subsection{High Level RTL Generation Frameworks}
High-level synthesis frameworks have emerged as powerful tools for generating optimized hardware implementations from high-level algorithmic specifications. However, the HLS design flow itself can still present a significant barrier to many users, as it requires a deep understanding of the underlying hardware architecture and optimization techniques. To address this challenge, a new generation of high-level RTL generation frameworks have been developed, which provide an additional layer of abstraction on top of traditional HLS tools. 

SODA~\cite{chi2018soda} and StencilFlow~\cite{de2021stencilflow} are examples of such frameworks that target the synthesis of stencil computations on FPGAs, abstracting away the low-level HLS details. HLS4ML~\cite{fahim2021hls4ml} is a toolkit that enables the deployment of machine learning models on FPGA accelerators using a high-level workflow. Other frameworks ~\cite{hegarty2014darkroom, wang2017comprehensive, canis2011legup} also allow developers to express their algorithms using domain-specific languages, which are then automatically transformed into optimized HLS code and ultimately synthesized into RTL. 


Although these frameworks successfully raise the level of abstraction, they are often tied to specific domains or demand substantial manual customization for new use cases. Moreover, their optimization techniques may fail to capture the complex patterns and trade-offs of modern hardware, limiting their ability to produce highly efficient implementations.

\section{Related Work}
The emergence of LLMs has spurred research on code generation and source-to-source translations~\cite{wang23survey_codegen}, including test-driven approaches that use prompts containing test cases to improve accuracy~\cite{tdd-codegen-Mathews, tdd-codegen-Fakhoury}. Open-source repositories provide abundant training data, enabling models to learn syntax, semantics, and optimization patterns, while recent agentic workflows enhance reliability~\cite{dong25survey_agent_codegen}.


LLM-based HDL generation has progressed from one-shot RTL synthesis to iterative workflows that generate, simulate, and repair Verilog using EDA feedback. Thakur \emph{et al.} introduced Verigen~\cite{Verigen2024}, showing that small fine-tuned models can outperform larger general-purpose LLMs in producing syntactically correct Verilog. Subsequent work incorporated iterative feedback and structure-aware methods, including RTLCoder~\cite{rtlcoder}, which outperforms GPT-4 on the VerilogEval Machine benchmark. Systems such as AutoChip~\cite{autochip}, RTLFixer~\cite{rtlfixer}, and VerilogCoder~\cite{verilogcoder} further improved correctness by leveraging compilation feedback, retrieval-augmented prompting, and multi-agent debugging, with VerilogCoder reporting up to $94.2$


Since FPGAs also support HLS, recent works explore LLMs for generating and optimizing HLS C/C++ code. Liao \emph{et al.}~\cite{survey_hls_llm} surveyed the field and highlight design challenges. Bhattacharyya \emph{et al.}~\cite{llmvshls2024} demonstrated LLM-driven HLS-to-RTL conversion with pragma insertion. HLSPilot~\cite{hlspilot} and C2HLSC~\cite{c2hlsc2024} automate C-to-HLS translation, profiling, and design space exploration, though context-length limits remain for large kernels. Complementary approaches include hierarchical preprocessing in C2HLSC and LLM-driven HLS program repair with retrieval and bit-width optimization by Xu \emph{et al.}~\cite{xurepair}.


We propose an LLM-driven agentic workflow, LAAFD, that automatically produces C\texttt{++} kernels tailored for Vitis HLS. Using C\texttt{++} instead of RTL as HLS descriptions yields more concise output, lowers token cost, and improves functional correctness. Compilation and co-simulation also provide rapid validation. Similar to prior work~\cite{c2hlsc2024}, our approach leverages HLS reports to guide the workflow, but we specifically target HPC kernels that involve deeply nested loops and exceed $250$ lines of code. In contrast to previous work, and in particular HLSPilot, LAAFD aggressively explores a large HLS optimization space with the explicit goal of minimizing kernel latency. To the best of our knowledge, this is the first tool capable of reaching the minimum theoretical latency across a wide range of challenging kernels. Our evaluation includes complex stencil kernels drawn from SODA~\cite{chi2018soda}, a class of benchmarks known for their difficulty in FPGA targeting, as evidenced by the numerous HLS-based methodologies~\cite{kamalakkannan2022fpga, Waidyasooriya2017, chi2018soda, de2021stencilflow, SpaceAndTemp} proposed for stencil applications. We evaluate LAAFD against manually tuned baselines and state-of-the-art stencil generators, and compare GPT-5 within this workflow against the GPT-4-mini model used in~\cite{c2hlsc2024}.

\section{Methodology}

\subsection{Agentic Workflow}
\begin{figure}[tbp]
  \centering
  \includegraphics[width=0.8\columnwidth]{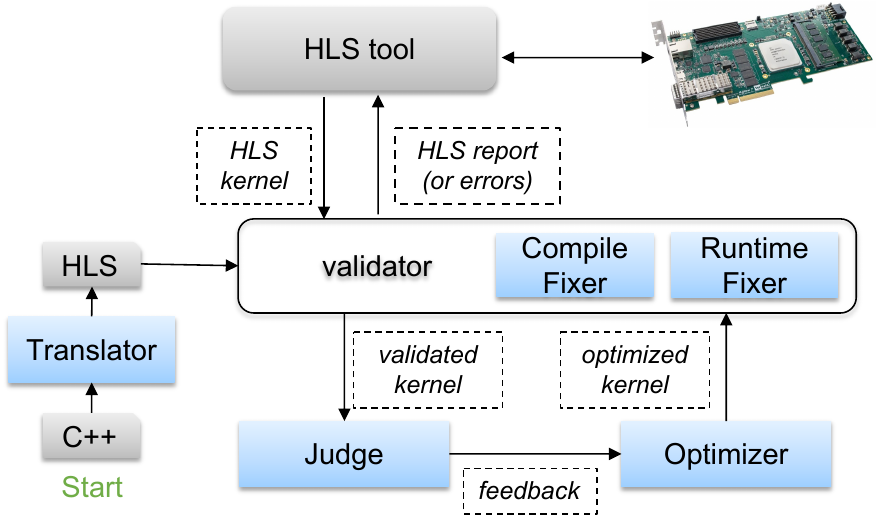}
  \caption{Simplified vision of the workflow for translating, compiling, validating, and optimizing pure C\texttt{++} code into C\texttt{++} HLS for Vitis.}
  \label{fig:workflow}
\end{figure}

The proposed workflow, shown in Figure~\ref{fig:workflow}, establishes an autonomous agent-based system that transforms standard C\texttt{++} kernels into optimized Vitis HLS code. The pipeline integrates translation, compilation, functional verification, and performance-driven optimization into a unified process. These phases are applied iteratively until the resulting kernel is both functionally correct and close to the theoretical performance bound.

\subsubsection{Translation}
This is the most straightforward step and is performed by a single agent. The agent receives the initial C\texttt{++} kernel along with the signature of the output kernel. Its task is to translate the input into HLS code. The agent may modify parameter types but must preserve both the number of parameters and the output kernel name. This constraint ensures interface consistency between the input and output kernels. At the same time, modifying parameter types is often essential for generating efficient HLS code. For instance, scalar parameters may need to be converted into fixed-point or arbitrary precision integer types to reduce hardware resource usage, or widened to enable vectorization and improve memory bandwidth utilization. Without such adjustments, it is generally impossible to obtain fully optimized HLS code, since the synthesis tool relies on type information to infer parallelism, pipeline structures, and efficient hardware data paths. Furthermore, the agent is instructed to insert loop trip counts and the required HLS include directives.

\subsubsection{Validation}
The validation phase ensures that the generated kernel both compiles and produces correct output. The process begins with compilation of the generated C\texttt{++} kernel. If compilation fails, the kernel together with the error messages is forwarded to the \texttt{compile fixer}, whose role is to correct compilation issues prior to HLS synthesis. Once compilation succeeds, we proceed to functional testing. For each kernel, we implemented a reference test that executes both the generated kernel and the original C\texttt{++} kernel on the same input, and then compares their outputs. If this validation step fails, the kernel, the validation error, and the test code are forwarded to the \texttt{runtime fixer}, whose role is to resolve functional mismatches without modifying the intended semantics of the kernel.

When both compilation and functional testing succeed, the process produces an HLS report that includes performance metrics required for subsequent phases, in particular latency and resource utilization. 

If validation exceeds a fixed time or iteration threshold, the workflow reverts to the most recent valid kernel. In the special case where no previous valid kernel exists (for example, during the translation stage), the workflow terminates and reports that the kernel cannot be translated. In practice, however, we have not observed such failures, even when using smaller LLMs.

\subsubsection{Optimization}
The central phase of the workflow is optimization, which is responsible for closing the gap between a functionally correct kernel and one that fully exploits the underlying hardware resources. This stage is often the most challenging in high-level synthesis, as it requires navigating a large design space of possible transformations while ensuring functional correctness. To address this complexity, the workflow adopts the \textit{LLM-as-a-judge} paradigm, where a language model is used not to generate new code directly but to act as an evaluator of existing implementations. In this role, the model examines the generated kernel and its HLS report against defined performance objectives, and produces structured feedback to guide further optimization. This functionality is encapsulated in the \texttt{judge} agent, described next.

After a kernel has been validated, it is passed to the \textit{judge}. It analyzes the kernel in conjunction with its HLS report and determines whether the implementation achieves the theoretical minimum performance. In such cases, the \texttt{judge} returns the decision \texttt{pass}. If, however, the kernel remains suboptimal (\emph{i.e.}, the latency can be further reduced), the \texttt{judge} issues structured feedback containing explicit optimization directives. If the \texttt{judge} returns \texttt{pass}, the workflow terminates immediately, since the kernel is deemed to have reached the theoretical performance bound and no further improvements are possible. Otherwise, the \texttt{judge}’s feedback is forwarded to the \texttt{optimizer} agent, which applies the proposed transformations and resubmits the modified kernel for validation. This forms an iterative optimization loop that proceeds until the \texttt{judge} issues a \texttt{pass} decision or until a predefined limit on the number of optimization iterations is reached.

Both the \texttt{judge} and the \texttt{optimizer} operate with access to the HLS report, which provides detailed performance metrics such as latency, initiation interval, and resource utilization. Additionally, each agent is executed within an independent context session. This design prevents uncontrolled growth of shared context, facilitates targeted reasoning, and preserves the modularity of agent interactions. It also reflects the standard prompt–response paradigm commonly used with large language models, in which agents are specialized and maintain isolated conversational states.

Additionally, the \texttt{judge} is also provided with a list of possible HLS optimizations. This list covers common techniques such as loop pipelining and vectorization, as well as a few more advanced options such as window buffering. 

\subsubsection{Illustrative Example}
To demonstrate the workflow in action, we walk through a complete example. This case study illustrates how the system takes a C\texttt{++} kernel, translates it into HLS code, incrementally transforms it, and finally converges to an implementation that achieves near-theoretical performance.

The process begins with the original C\texttt{++} kernel shown in Figure~\ref{fig:ex-original}. This kernel applies a row-wise vector addition: each row of the two-dimensional array \texttt{A} is incremented by the corresponding scalar from array \texttt{B}. This kernel is provided as input to the \texttt{translator}, which generates the HLS version in Figure~\ref{fig:ex-translated}. The translation introduces \texttt{extern "C"} interfaces and \texttt{LOOP\_TRIPCOUNT} pragmas, ensuring compatibility with Vitis HLS. However, the generated code fails to compile because the type \texttt{vec\_type} is undefined. This triggers the validation phase of the workflow. The validator detects the compilation error and forwards the kernel to the \texttt{compile fixer}. By introducing the missing type definition, the agent produces the corrected version shown in Figure~\ref{fig:ex-fixed}. This code compiles successfully and passes functional testing. However, the nested loops still execute sequentially, leading to a latency of about 65{,}543 cycles, which is far from optimal.

\begin{figure}[tbp]
  \centering
  \begin{minipage}{0.98\columnwidth}
\begin{lstlisting}[morekeywords={pragma,HLS}]
#define P_SIZE 256
void golden_model(std::vector<int> A,
                  std::vector<int> B,
                  std::vector<int> &result){
  for(int i = 0; i < P_SIZE; i++)
    for(int j = 0; j < P_SIZE; j++)
      result[i*P_SIZE+j] = A[i*P_SIZE+j] + B[i];
}
\end{lstlisting}
  \end{minipage}
  \caption{Initial C\texttt{++} kernel provided as input to the workflow.}
  \label{fig:ex-original}
\end{figure}

\begin{figure}[tbp]
  \centering
  \begin{minipage}{0.98\columnwidth}
\begin{lstlisting}[morekeywords={pragma,HLS}]
#define P_SIZE 256
extern "C" {
void genKernel(const vec_type *A,
               const vec_type *B,
               vec_type *C) {
  #pragma HLS LOOP_TRIPCOUNT min=P_SIZE max=P_SIZE avg=P_SIZE
  for(int i = 0; i < P_SIZE; i++) {
    #pragma HLS LOOP_TRIPCOUNT min=P_SIZE max=P_SIZE avg=P_SIZE
    for(int j = 0; j < P_SIZE; j++)
      C[i*P_SIZE + j] = A[i*P_SIZE + j] + B[i];
  }
}}
\end{lstlisting}
  \end{minipage}
  \caption{Initial code translation produced by the \texttt{translator} agent.}
  \label{fig:ex-translated}
\end{figure}

\begin{figure}[tbp]
  \centering
  \begin{minipage}{0.98\columnwidth}
\begin{lstlisting}
#define P_SIZE 256
typedef int vec_type;
extern "C" {
void genKernel(const vec_type *A,
               const vec_type *B,
               vec_type *C) {
  #pragma HLS LOOP_TRIPCOUNT min=P_SIZE max=P_SIZE
  for(int i = 0; i < P_SIZE; i++) {
    #pragma HLS LOOP_TRIPCOUNT min=P_SIZE max=P_SIZE
    for(int j = 0; j < P_SIZE; j++)
      C[i*P_SIZE + j] = A[i*P_SIZE + j] + B[i];
  }
}}
\end{lstlisting}
  \end{minipage}
  \caption{Corrected kernel produced by the \texttt{compile fixer} agent.}
  \label{fig:ex-fixed}
\end{figure}

The kernel is now valid but not yet efficient. The workflow advances to the next stage: the \texttt{judge} analyzes the implementation together with its HLS report. This evaluation provides structured feedback on how the design can be improved. The feedback, shown in Figure~\ref{fig:ex-judge1}, identifies several missed optimization opportunities, including loop pipelining, vectorization, and dataflow optimization.

\begin{figure}[tbp]
  \centering
  \begin{minipage}{0.98\columnwidth}
  \begin{tcolorbox}[colback=gray!4,colframe=black!30,boxrule=0.4pt,sharp corners]
  \footnotesize
The current kernel runs the 256$\times$256 loop fully sequentially ($\sim$65{,}543 cycles), so there is still headroom. To slash the latency you should:
\begin{enumerate}\itemsep2pt
  \item Hoist \texttt{B[i]} load out of the inner j-loop so you only read B once per row.
  \item Pipeline the inner loop with II=1 (\#pragma HLS pipeline II=1).
  \item Vectorize by a factor of 4 on a 128-bit data path: pack 4 values into one AXI transfer, unroll the inner loop by 4, and operate on 4 elements/cycle to get a 4× speedup.
  \item (Optional) Separate memory-load, compute and write into three functions and apply \#pragma HLS dataflow to overlap A/B reads, compute, and C writes.
\end{enumerate}
With these changes you can reduce the cycles from $\sim$65{,}536 to $\sim$16{,}384 (plus a small pipeline fill/drain), cutting your latency by 4×.
  \end{tcolorbox}
  \end{minipage}
  \caption{Feedback produced by the \texttt{judge} agent highlighting optimization opportunities.}
  \label{fig:ex-judge1}
\end{figure}

Guided by this structured feedback, the \texttt{optimizer} introduces transformations such as vectorization or loop pipelining. Each candidate kernel is validated, and the process iterates until a stable solution emerges. After several iterations, the workflow converges to the optimized kernel skeleton in Figure~\ref{fig:ex-optimized}. The full implementation is considerably longer, but for clarity only the structure is shown here.

\begin{figure}[tbp]
  \centering
  \begin{minipage}{0.98\columnwidth}
\begin{lstlisting}[morekeywords={pragma,HLS}]
#define P_SIZE 8192
#define VEC_SIZE 4
typedef ap_uint<VEC_SIZE*32> vec_type;

extern "C" void genKernel(const vec_type *A,
                          const vec_type *B,
                          vec_type *C) {
  #pragma HLS DATAFLOW
  read_data_A(A, sA);
  read_data_B(B, sB);
  compute(sA, sB, sC);
  write_data(C, sC);
}
\end{lstlisting}
  \end{minipage}
  \caption{Skeleton of the final optimized kernel.\\ (details omitted for brevity)}
  \label{fig:ex-optimized}
\end{figure}

The validator confirms that this optimized kernel is both correct and synthesizable. Finally, the \texttt{judge} approves the design with the feedback shown in Figure~\ref{fig:ex-judge2}, certifying that the implementation achieves a latency of \(\approx16{,}396\) cycles, which is essentially the theoretical minimum of 16{,}384 iterations plus pipeline overhead. At this point, the workflow terminates.

\begin{figure}[tbp]
  \centering
  \begin{minipage}{0.98\columnwidth}
  \begin{tcolorbox}[colback=gray!4,colframe=black!30,boxrule=0.4pt,sharp corners]
  \footnotesize
The kernel already achieves a fully overlapped dataflow with II=1 on all stages, uses 128-bit AXI for A/C, buffers B on-chip, and vectorizes the 4-float tiles. Its throughput (\(\approx16{,}396\))  is essentially the theoretical minimum for 16,384 tiles plus pipeline overhead. No further latency reduction is possible.
  \end{tcolorbox}
  \end{minipage}
  \caption{Final \texttt{judge} feedback certifying optimal performance.}
  \label{fig:ex-judge2}
\end{figure}

In summary, this example illustrates the complete workflow: the \texttt{translator} generates an initial HLS kernel, the validator ensures correctness, the \texttt{judge} provides targeted optimization guidance, and the \texttt{optimizer} applies transformations until convergence. Together, the agents translate a C\texttt{++} implementation into optimized HLS code that reaches the practical performance limit.

\subsection{Kernels and HLS Optimizations}

\begin{table*}[hbt]
  \caption{Kernels description}
  \rowcolors{2}{gray!25}{white}
  \label{tab:Kernels}
  \centering
  \begin{tabular}{llllll} 
    \toprule
    Kernel & Description & dimension & ideal (cycles) & manual  & LAAFD  \\
    \midrule
    MX & Maximum value in an integer array & $8192$ & 2,048 & 2,073 & 2,062 \\
    AMX & Index of of the maximum value in an integer array & $8192$ & 2,048 & 2,073 &  2,061\\
    AXPY & Integer computation of $\alpha \times X[i] + Y[i] $ & $8192$ &  2,048 & 2,069 & 2,069 \\
    CSUM & Summation along column, $C[i] = sum(A[:,i])$ & $1024 \times 1024$ &  262,144 & 262,675 &  262,195\\
    LF & Fuse two loops and eliminate intermediate storage & $8192$ &  2,048 & 2,069 &  2,067 \\ 
    SREP & $A[i,j] = B[i,j] + C[i,i]$ & $256 \times 256$ &  16,384 & 16,404 & 16,407\\
    VADD & Vector add operation & $8192$ &  2,048 & 2,068 & 2,067\\
    S2D & Four point 2-D stencil operation & $1024 \times 1024$ & 262,400 & 262,413 &  262,421\\
    S2D2 & Two iteration of four point 2-D stencil operation & $1024 \times 1024$ &  262,656 & 262,416 & 262,210 \\
    S2DB & Batched computation of four point 2-D stencil operation & $1024 \times 1024 \times 4$ &  1,048,832  & 1,048,845 & $1,049,702$\\
    S2DN & N=4 iteration of four point 2-D stencil operation & $1024 \times 1024$ &  263,168 & 262,422 &  262,432  \\ 
    S2NB & N=4 iteration of Batched four point 2-D stencil operation & $1024 \times 1024 \times 8$ &  2,097,408 & 2,097,430  & 2,105,364 \\ 
    S3D & Six point 3-D stencil operation & $256 \times 256 \times 256$ &  4,210,688 &  4,210,703 & 4,210,846   \\
    S3DB & Batched computation of six point 3-D stencil operation & $256 \times 256 \times 256 \times 4$ &  16,793,600 & 16,793,615 & 16,874,425 \\
    VDOT & Vector dot operation & $8192$ & 2,048 &  2,063  &  2,063\\
    \bottomrule
  \end{tabular}
\end{table*}

\begin{table*}[ht]
  \caption{Kernels and HLS optimizations}
  \label{tab:kernel-opt}
  \centering
  \footnotesize
  \setlength{\tabcolsep}{3pt} 
  \rowcolors{2}{gray!25}{white}
  \begin{tabular}{llllllllllllllll} 
    \toprule
    Optimizations(Memory (M), Compute(C))  & MX  & AMX & AXPY & CSUM & LF & SREP & VADD & S2D & S2D2 & S2DB & S2DN & S2NB & S3D & S3DB & VDOT \\
    \midrule
    Pipelining (C) & $\checkmark$ & $\checkmark$ & $\checkmark$ & $\checkmark$ & $\checkmark$ & $\checkmark$ & $\checkmark$ & $\checkmark$ & $\checkmark$ & $\checkmark$ & $\checkmark$ & $\checkmark$ &$\checkmark$ &  $\checkmark$ &  $\checkmark$  \\
    Vectorization (C,M) & $\checkmark$ & $\checkmark$ & $\checkmark$ & $\checkmark$ & $\checkmark$ & $\checkmark$ & $\checkmark$ & $\checkmark$ & $\checkmark$ & $\checkmark$ & $\checkmark$ & $\checkmark$ &$\checkmark$ &$\checkmark$ &  $\checkmark$ \\
    Modularized and streaming (M) & $\checkmark$ & $\checkmark$ & $\checkmark$ & $\checkmark$ & $\checkmark$ & $\checkmark$ & $\checkmark$ & $\checkmark$ & $\checkmark$ & $\checkmark$ & $\checkmark$ & $\checkmark$ &$\checkmark$ &$\checkmark$ &  $\checkmark$ \\
    Dataflow (C) & $\checkmark$ & $\checkmark$ & $\checkmark$ & $\checkmark$ & $\checkmark$ & $\checkmark$ & $\checkmark$ & $\checkmark$ & $\checkmark$ & $\checkmark$ & $\checkmark$ & $\checkmark$ &$\checkmark$ &$\checkmark$ &  $\checkmark$  \\
    Loop flattening (C) & & & & $\checkmark$ & & & & $\checkmark$ & $\checkmark$ & $\checkmark$& $\checkmark$ & $\checkmark$ & $\checkmark$ & $\checkmark$  \\
    Loop-invariant code motion (M) & & & & & &$\checkmark$  &  &   &  &  & & & & & \\
    Localized reduction (C, M) & $\checkmark$ & $\checkmark$ & & & & & & & & & & & &  \\
    Perfect data reuse (M) & & & & & & & & $\checkmark$ & $\checkmark$ & $\checkmark$ & $\checkmark$ & $\checkmark$ & $\checkmark$ & $\checkmark$ &  \\
    coalesced memory access (M) & & & &  $\checkmark$  &  \\ 
    Loop fusion (C, M) & & & & &  $\checkmark$ & & & & & & & & & & \\
    Unroll iterative loop (C) & & & & & & & & & & & &  $\checkmark$  & &  $\checkmark$  &  \\
    Batching and loop reordering (C) & & & & & & & & & &  $\checkmark$  & &  $\checkmark$  & &  $\checkmark$  & \\
    Intermediate data structure removal (M) & & & & & $\checkmark$ & & & & $\checkmark$ & & $\checkmark$ & & & \\
    shift buffers (C) & & & & & & & & $\checkmark$ & $\checkmark$ & $\checkmark$ & $\checkmark$ & $\checkmark$ & $\checkmark$ & $\checkmark$ &  \\
    \bottomrule
  \end{tabular}
\end{table*}

To evaluate LAAFD, we have designed a set of kernels with varying levels of complexity, ranging from simple single-loop structures to deeply nested loops involving multiple memory access operations. These kernels are intended to evaluate both fundamental HLS optimizations such as pipelining, vectorization, dataflow, and advanced techniques, including perfect data reuse and iterative loop unrolling for stencil-based applications. No further implementation details are embedded within the kernels themselves. The complete list of stencil kernels used in our experiments is presented in Table~\ref{tab:Kernels}. For each kernel, a set of targeted optimizations was applied, as summarized in Table~\ref{tab:kernel-opt}. 

We also take the set of kernels listed in Table~\ref{tab:Soda_Kernels} from SODA\cite{chi2018soda}, in order to compare the generated code using LAAFD with the state of the art of stencil specific HLS code generator, SODA. These include 2D and 3D kernels representing both multi-stage and multi time step stencil operations. These kernels are specified using SODA DSL in~\cite{chi2018soda} and we wrote equivalent C++ golden models for each kernel. LAAFD takes this C++ golden model and generates the Vitis HLS kernel. Here we note that the SODA generated kernel requires specific data layout arrangement for the inputs as well as outputs, but kernels generated using LAAFD don't require any data layout transformations.

\section{Evaluation}
In this section, we evaluate the kernels produced by LAAFD against both manually optimized kernels and those generated by the state-of-the-art SODA framework~\cite{chi2018soda} for stencil solvers. The hardware configuration is fixed to a 128-bit vector path with a loop unroll factor of four for stencil kernels. Our comparison emphasizes execution cycles, FPGA area utilization, and the readability and maintainability of the generated code. As summarized in Table~\ref{tab:ExSetup}, LAAFD leverages LLMs such as GPT-5, GPT-5-nano, and o4-mini to construct its agents. The kernels are synthesized using the Vitis 2022.2 HLS tool and targeted for implementation on the xcu250-figd2104-2L-e FPGA at a $200$ MHz frequency. All FPGA implemented kernels operate with $128$-bit input and output data paths. For this evaluation, we assume that the external memory bandwidth can sustain multiple $128$-bit input and output transfers at the $200$ MHz clock rate.   

\begin{table}[ht]
  \caption{Experimental setup}
  \label{tab:ExSetup}
  \centering
  \small
  \rowcolors{2}{gray!25}{white}
  \begin{tabular}{ll} 
    \toprule
    Kernel & Description  \\
    \midrule
    HLS tool & Vitis 2022.2  \\
    FPGA device & xcu250-figd2104-2L-e  \\
    Target Frequency & 200 MHz \\
    LLMs for the Agents & gpt-5, 04-mini, gpt5-nano \\
    \bottomrule
  \end{tabular}
\end{table}

\subsection{Execution cycles}
To evaluate the runtime performance of the generated kernels against the baselines, we use the cycle counts reported by the HLS tool. Table~\ref{tab:Kernels} presents the theoretical minimum cycle (\texttt{ideal}) counts alongside results for both manually tuned baselines and LAAFD-generated kernels. Our results show that LAAFD-generated kernels achieve performance comparable to manually optimized baselines. Overall, LAAFD attains $99.9$\% of the geometric mean performance of the hand-tuned baselines when using a state-of-the-art model, such as GPT-5 for kernels in Table~\ref{tab:Kernels}. As noted in Section~\ref{sec:HLS}, HLS tools do not account for dataflow delays between modules when estimating total latency. Consequently, in some cases, the cycle counts of LAAFD generated kernels and the manually tuned baseline are reported to be lower than the \texttt{ideal} count.

Table~\ref{tab:Soda_Kernels} presents a comparison of the execution cycles of kernels generated by SODA against the \texttt{ideal} latency. Since SODA does not insert trip count pragmas for the \texttt{for} loops in its generated kernels and because these kernels contain many such loops, we evaluate the execution time of LAAFD generated kernels relative to \texttt{ideal} using the formula described in prior works~\cite{kamalakkannan2022fpga, chi2018soda}. The kernels from SODA~\cite{chi2018soda} are particularly challenging, as they include multiple nested loops and their high-level C\texttt{++} golden models can reach up to $250$ lines of code. Despite this complexity, LAAFD successfully generated equivalent Vitis HLS kernels that achieve latency comparable to \texttt{ideal}. It is worth noting that SODA employs its own DSL for describing stencil-like kernels, whereas LAAFD accepts high-level C\texttt{++} specifications for arbitrary kernel types.

We also evaluated the estimated execution cycles obtained by the best implementation of each of the three language models we tested across the $15$ kernels defined in Table~\ref{tab:Kernels}. 



\begin{figure}[tbp]
    \centering
    \includegraphics[width=\columnwidth]{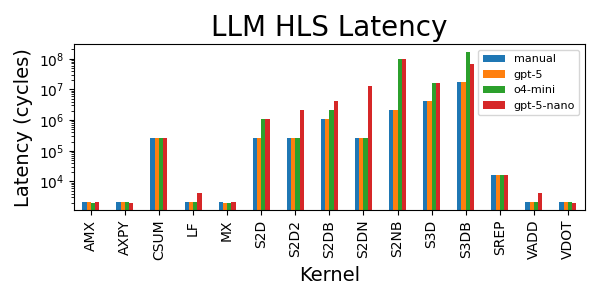}
    \caption{HLS implementation latency comparison between an optimized references implementation and LLM generated and optimized implementations.}
    \label{fig:absolute-results}
\end{figure}




\begin{table*}[ht]
  \caption{Kernels from SODA : L - LAAFD generated, S - SODA generated}
  \label{tab:Soda_Kernels}
  \centering
  \rowcolors{2}{gray!25}{white}
  \begin{tabular}{llllllllll} 
    \toprule
    Kernel & Dim & LoC(L/S) & LUT(L/S) & FF(L/S) & DSP(L/S) & BRAM(L/S) & Ideal  & LAAFD  \\
    \midrule
    blur & $1024\times 1024$ & $196/1,330$ & $23,944/29,149$ & $18,036/16,604$ & $96/96$ & $68/40$ &  264,192 & 262,717\\
    contrast & $1024\times 1024$ & $255/7,800$ & $243,688/371,156$ & $351,064/457,888$ & $3,816/3,932$ & $88/112$ &  278,528 & 262,604\\
    denoise2d & $1024 \times 1024$ & $410/1,829$ & $45,487/58,850$ & $50,387/64,622$ & $414/424$ & $76/20$ & 262,656 & 264,077\\
    erosion & $1024 \times 1024$ & $233/1,988$ &  $26,778/47,911$ & $7,869/26,668$ & $0/0$ & $84/90$ & 266,752 & 267,276\\ 
    heat3d & $256\times 256 \times 256$ & $250/1,625$ & $51,967/66,320$ & $60,999/72,419$ & $316/528$ & 2,756/544 & 4,259,840 & 4,238,313\\
    jacobi2d & $1024 \times 1024$ & $160/1,229$ & $31309/41137$ & $30838/39414$ & $176/176$ & $116/32$ &  263,168 & 262,554  \\
    seidel2d & $1024 \times 1024$ & $260/1,726$ & $79512/63159$ & $102123/63016$ & $304/304$ & $344/32$ & 263,168 & 262,246\\
    \bottomrule
  \end{tabular}
\end{table*}

\subsection{Consumed FPGA resources}
The resource utilization comparison between LAAFD-generated kernels and manually tuned baseline/SODA-generated kernels is reported in Table~\ref{tab:FPGA_res} and Table~\ref{tab:Soda_Kernels}, respectively. As LAAFD primarily focuses on reducing cycle counts, it generally incurs higher resource usage for the given problem size, as shown in Table~\ref{tab:FPGA_res}. This overhead arises from several factors. First, LAAFD introduces deeper \texttt{hls::stream} buffers, which require block memories, whereas the manually tuned baseline and SODA generated kernels achieve similar functionality with shallower streams implemented using shift registers. In some cases, LAAFD modularizes the kernel deeply and connects with \texttt{hls::streams} with higher depth, leading to more BRAM usage. Second, LAAFD incorporates additional logic to handle corner cases (\emph{e.g.}, problem sizes not being multiples of the vectorization factor), even though such cases do not occur in the evaluated workloads. In the \texttt{CSUM} kernel, for instance, LAAFD applies tiling optimizations that would only be effective for much larger problem sizes, leading to unnecessary resource consumption at smaller scales.

The overhead is particularly pronounced in stencil kernels, where efficient data reuse requires more complex buffer architectures. Rather than adopting the window-buffer approach proposed in prior works~\cite{kamalakkannan2022fpga, chi2018soda}, LAAFD employs cyclically partitioned memory banks, which significantly increases BRAM usage (\emph{e.g.}, $576$ blocks in \texttt{S3D}). For the evaluated problem size of $256$, each BRAM block is only partially utilized, making this approach inefficient for smaller problem sizes. In Table~\ref{tab:Soda_Kernels}, we can see a similar pattern where LAAFD generated kernels consume more BRAM resources than SODA generated kernels while also consuming a similar number of DSP resources. 

It should be noted that the workflow was not instrumented to optimize resource utilization at any stage. Our focus was exclusively on minimizing latency. Indeed, for the SODA kernels, we even had to remove information about resource utilization from the HLS report provided to the \texttt{judge} agent, in order to remain within the context size limits. We expect that more favorable trade-offs between latency and resource usage could be obtained by adapting the prompting strategy given to the agents.

\begin{table}
  \caption{FPGA Resource consumption: LAAFD/Manual}
  \label{tab:FPGA_res}
  \centering
  \footnotesize
  \setlength{\tabcolsep}{3pt} 
  \rowcolors{2}{gray!25}{white}
  \begin{tabular}{llllll} 
    \toprule
    Kernel & LoC & LUT & FF & DSP & BRAM  \\
    \midrule
    MX & $57/66$ & $5,916/3,942$ &  $3,191/2,451$ & $0/0$ & $4/0$ \\
    AMX & $64/74$ & $5,255/4295$ &  $3483/2654$ & $0/0$ & $0/0$ \\
    AXPY & $79/75$ & $12,481/10,244$ &  $6,761/6,634$ & $12/12$ & $8/0$  \\
    CSUM & $119/82$ & $20,335/8,913$ &  $21,000/4,377$ & $0/0$ & $8/12$\\
    LF & $42/75$ & $8,975/10,244$ &  $5,445/6,634$ & $12/12$ & $0/0$  \\ 
    SREP & $97/94$ & $13,192/10,264$ &  $6,414/6,371$ & $0/0$& $0/0$ \\
    VADD & $87/73$ & $12,527/10,007$ &  $6,468/6,169$ & $0/0$ &  $0/0$ \\
    S2D & $164/124$ & $6,350/9,648$ &  $4,746/5,209$ & $0/0$ & $8/16$ \\
    S2D2 & $299/130$ & $34,306 /11,477$ &  $28,566/6,681$ & $0/0$ & $90/28$  \\
    S2DB & $173/126$ & $10,989/9,686$ &  $6,042/5,209$  & $0/0$ & $32/16$ \\
    S2DN & $180/130$ & $18,139/14,743$ &  $12,583/9,411$ & $0/0$ & $84/52$   \\ 
    S2NB & $170/134$ & $15,910/14,885$ &  $9,261/9,405$ & $0/0$  & $56/52$  \\ 
    S3D & $156/172$ & $29,840/11,072$ & $17,426/7,506$  &  $18/0$ & $584/122$    \\
    S3DB & $240/172$ & $4,528/10,942$ &  $4,294/7,253$ & $1/0$ & $521/122$  \\
    VDOT & $89/72$ & $11,923/7,194$ & $6,721/4,739$ &  $12/12$  & $0/0$\\
    \bottomrule
  \end{tabular}
\end{table}


\subsection{Code quality}
As a proxy to evaluate the code quality and complexity of the hand-tuned and generated kernels, we used lines of code (LoC) and estimated cyclomatic complexity calculated from the \textbf{S}loc \textbf{C}loc and \textbf{C}ode (scc) tool version 3.2~\cite{scc}. First, we compared the code quality between the manually hand-tuned code and the LAAFD generated code (GPT-5) for our suite of 15 kernels. From Table~\ref{tab:FPGA_res}, we can see that for $11$ of the $15$ kernels, LAAFD generated more LoC than the hand-tuned kernels, and on average contain $1.16\times$ more lines per HLS kernel than the hand-tuned code. However, the cyclomatic complexity actually decreased on average, where each LAAFD generated kernel had $0.89\times$ the complexity of a hand-tuned kernel (Table~\ref{tab:codequality}). This demonstrates that LAAFD is able to generate code that is near or on par with the quality of code in a hand-tuned kernel while also achieving similar performance.

Next, we compared the code quality of the generated LAAFD code and the SODA HLS for the $7$ SODA kernels evaluated. SODA is a complex DSL which generates optimized, but long and complex HLS code that can be difficult to read. As seen in Table~\ref{tab:Soda_Kernels}, the SODA code required $4.5\times$ - $30\times$ more lines of code compared to the equivalent LAAFD generated code. Furthermore, SODA required $8.3\times$ more lines of code per kernel with $2.27\times$ more cyclomatic complexity, on average, than the LAAFD generated code, demonstrating that LAAFD can generate near optimal HLS that is also high quality, readable, and less complex than the equivalent SODA implementation.

\begin{table}[ht]
  \caption{Code Complexity: S - SODA, L - LAAFD, M - Manual}
  \label{tab:codequality}
  \centering
  \footnotesize
  \rowcolors{2}{gray!25}{white}
  \begin{tabular}{llllllll} 
    \toprule
    Kernel & \makecell{Complexity\\(L/M)} & Ratio & & Kernel & \makecell{Complexity\\(S/L)} & Ratio \\
    \midrule
    MX & 3/7 & 0.43 & & blur & 31/22 & 1.41 \\
    AMX &7/9 & 0.78 & & contrast & 481/28 & 17.18 \\
    AXPY & 5/5 & 1.00 & & denoise2d & 73/74 & 0.99 \\
    CSUM & 12/9 & 1.33 & & erosion & 73/30 & 2.43 \\
    LF & 1/5 & 0.20 & & heat3d & 45/31 & 1.45 \\
    SREP & 10/15 & 0.67 & & jacobi2d & 39/20 & 1.95 \\
    VADD & 5/5 & 1.00 & & seidel2d & 51/27 & 1.89 \\
    S2D & 26/25 & 1.04 & & & & \\
    S2D2 & 64/25 & 2.56 & & & & \\
    S2DB & 51/29 & 1.76 & & & & \\
    S2DN & 38/26 & 1.46 & & & & \\
    S2NB & 25/30 & 0.83 & & & & \\
    S3D & 27/36 & 0.75 & & & & \\
    S3DB & 38/35 & 1.09 & & & & \\
    VDOT & 4/6 & 0.67 & & & & \\
    GeoMean & & 0.89 & & & & 2.27 \\
    \bottomrule
  \end{tabular}
\end{table}

\subsection{Comparison of Models}
Across the 3 LLMs tested, we evaluated the estimated latency of the LAAFD generated HLS across our $15$ kernel suite to compare their overall performance and trends (Figure~\ref{fig:absolute-results}). Each LLM was able to achieve higher than $98$\% of the ideal and manual hand-tuned performance for several of the simpler kernels, such as MX, AMX, AXPY, CSUM, SREP, and VDOT. Specifically, GPT-5-nano achieved greater than $98$\% performance on $6$ of the $15$ kernels, while this was achieved on $10$ of the $15$ kernels for o4-mini, and for all $15$ kernels for GPT-5. 

While GPT-5 achieved high performance for all kernels, demonstrating the ability to apply all of the optimizations necessary to approach ideal and hand-tuned performance (Table~\ref{tab:kernel-opt}), GPT-5-nano and o4-mini achieved less than $50$\% optimal performance on the remaining $9$ and $5$ kernels, respectively. GPT-5-nano was able to approach ideal performance for the LF kernel and the 2 stencil kernels, all of which required intermediate data structure removal (S2D2, S2DN), while o4-mini did not appear to be able to successfully optimize the kernels which required this optimization. Both GPT-5-nano and o4-mini struggled on the remaining stencil kernels, all of which required perfect data reuse and shift buffer optimizations, indicating that a larger and more advanced model, such as GPT-5, is likely necessary to perform more complex optimizations such as these.

Overall, GPT-5 achieved a geomean performance across our suite of 15 HPC kernels of $99.9$\% compared to the hand-tuned baseline, while o4-mini achieved a geomean performance $52.5$\% and GPT-5-nano achieved a performance of $32.7$\%. 




\subsection{Discussions}

In this work, we report for each kernel the best HLS design obtained across multiple executions of the workflow. Each execution involves several iterations of translation, validation, and optimization, and different runs may yield different results even under identical settings, reflecting the stochastic nature of LLM-based agents. For the more complex SODA kernels, we ran the workflow ten times per kernel with $25$ optimization iterations each. Only one or two runs produced HLS code deemed optimal by the \texttt{judge} agent, with latency within $1$\% of the theoretical minimum; the remaining runs produced functionally correct but suboptimal designs within the given iteration budget. In contrast, simpler kernels such as \texttt{vdot}, \texttt{max}, and \texttt{amax} typically required only a few iterations to reach optimized, functional HLS code, sometimes even when using a smaller model such as GPT-5-nano. This suggests that the number of iterations required to reach optimal performance grows substantially with kernel complexity.


Scalability posed additional challenges for larger kernels, particularly in the SODA set, where executions sometimes exceeded the maximum context size. We addressed this by summarizing HLS reports or, in extreme cases, removing the dedicated \texttt{judge} session. While this limitation did not affect smaller kernels, it may hinder application to full-scale programs without improved methods for handling large reports.

Overall, the workflow can translate pure C\texttt{++} kernels into near-optimal HLS implementations, but its effectiveness depends on kernel complexity, semantic correctness, and context limitations.

\section{Conclusion}
We have demonstrated that LLM-based agents can effectively translate general-purpose C\texttt{++} into optimized HLS code. This approach has the potential to broaden adoption of FPGAs in both scientific and edge computing settings. In our study, $15$ representative HPC kernels were translated, achieving $99.9$\% of the performance of hand-tuned HLS kernels. For stencil workloads, the results matched the performance of the state-of-the-art SODA HLS code generator. Considering the relatively small amount of FPGA-related source code available on GitHub, which likely influenced LLM training, these results are especially noteworthy. Moreover, the cost of approximately US\$50 to translate and optimize all $15$ kernels highlights the feasibility of this approach, though server infrastructure costs are not included.

We compared three ChatGPT models: GPT-5-Nano, GPT-5, and o4-mini. GPT-5-Nano is a faster and more cost-efficient variant of GPT-5, achieved by trading off some reasoning depth and accuracy. As a result, its outputs were generally less precise than those of the other two models. o4-mini, from the previous generation, is also optimized for speed and affordability, but it is weaker than any GPT-5 variant in terms of reasoning, accuracy, and multimodal support. In contrast, GPT-5 is the flagship model, offering the highest reasoning quality and accuracy. 

\subsection*{Outlook}
This study focused on kernels; however, an important next step is to extend the methodology to full applications. Translating entire applications, rather than isolated kernels, would enable a more comprehensive evaluation of performance, scalability, and portability across diverse computational workloads.

Another promising direction lies in the fine-tuning of models using a larger and more diverse set of HLS examples. The current models may be constrained by training on relatively small or narrow datasets. Broadening the training corpus could strengthen the models’ generalization capabilities across varied HLS tasks. Such improvements may reduce the need for extensive manual optimizations and thereby lower overall development costs.

In addition, while this study concentrated on OpenAI models, there exist both closed-source and open-source LLMs that warrant investigation. Exploring the performance of different LLMs on this task could provide valuable insights into their relative strengths and limitations. A central challenge, however, is to establish a fair basis for comparison across models, accounting for factors such as computational cost, number of parameters, training data, and accuracy.

Ultimately, combining full-application translation, model fine-tuning, and cross-model evaluation has the potential to substantially advance the efficiency, accuracy, and practicality of HLS-based design workflows.




\section*{Acknowledgment}
This document is released by Los Alamos National Laboratory under LA-UR-26-20594 and has been approved for public release; distribution is unlimited. Los Alamos National Laboratory is operated by Triad National Security, LLC, for the National Nuclear Security Administration of the U.S. Department of Energy under contract 89233218CNA000001.

\newpage

\bibliographystyle{IEEEtran}
\bibliography{refs}

\end{document}